\documentstyle[preprint,revtex]{aps} 
\newcommand{\bea}{\begin{eqnarray}}
\newcommand{\eea}{\end{eqnarray}}
\newcommand{\beq}{\begin{equation}}
\newcommand{\eeq}{\end{equation}}
\newcommand{\nn}{\nonumber}
\newcommand{\st}{\scriptstyle}
\newcommand{\dst}{\displaystyle}
\tightenlines
\begin{document}
\begin{title}
Ramsey fringes in atomic interferometry: measurability of the
influence of space-time curvature
\end{title}
\author{J\"urgen Audretsch and Karl-Peter Marzlin}
\begin{instit}
Fakult\"at f\"ur Physik\\
der Universit\"at Konstanz\\
Postfach 5560 M674\\
D-78434 Konstanz, Germany
\cite{e-mail}
\end{instit}
\begin{abstract}
The influence of space-time curvature on quantum matter
which can be
theoretically described by covariant wave equations has not been
experimentally established yet. In this paper we analyze in
detail the suitability of the Ramsey
atom beam interferometer for the
measurement of the phase shift caused by the Riemannian
curvature of the earth or alternatively of two lead blocks.
It appears that for the lead blocks
the detection should be possible with realistic
modifications of existing devices within the near future.
For the earth's gravitational field the experimental
difficulties are too big.
The paper is divided into two parts. The first one is concerned
with the derivation of general relativistic correction terms
to the Pauli equation starting from the fully covariant Dirac
equation and their physical interpretation. The inertial effects
of acceleration and rotation are included. The calculation
makes use of Fermi coordinates.
In the second part we calculate the shift of the Ramsey
fringes for the two different sources of curvature and examine
various possibilities to enlarge the sensitivity of the
apparatus to space-time curvature.
Since the two parts may be more or less interesting for
physicists with different research fields they are written in
such a way that each one may be read without much reference to
the other one.
\end{abstract} $ $
\narrowtext
\section{Introduction}
The connection between general relativity and quantum theory
has been the subject of intense theoretical investigations
for decades. Nevertheless, due to the smallness of the influence
of gravity on quantum systems in the laboratory, there is
presently still a huge gap between the top theoretical level
of quantized gravity on one hand and the level of
empirical verification on the other. The latter is the subject

of this paper. If we neglect any quantum effects from gravity
itself as they are predicted by, e.g., quantum gravity or
superstring theory, we have to formulate quantum theory and
the laws of physics in general for arbitrarily moving observers
in a given curved space-time and to look for empirical
implications. Based on the results of Special
Relativity this formulation is done
by means of several theoretical principles like Einstein's
equivalence principle for example. The results are empirically
well confirmed for classical matter like test particles

and light rays. For quantum systems on  the other hand special
relativistic effects of course are demonstrated
up to extremely high energies. But up to recent years there were
practically no experiments establishing how quantum systems
react on gravity and inertia in genuine quantum effects.

This situation changed when matter wave interferometry with
electrons, neutrons, and especially atoms took advantage of
technological progress in the development of, e.g., single
crystals or lasers. By the study of the induced phase shift of
the fringe pattern it was possible to establish the influence of
non-inertial motion and of the homogeneous earth gravitational
field on quantum systems: Colella {\em et al.}
\cite{COW} measured the effect of the earth's acceleration
on neutrons. In contrast, Bonse and Wroblewski \cite{bonse83}

have demonstrated the influence of the constant acceleration
of the reference frame thus
showing the equality of inertial and gravitational mass for
neutrons (restricted version of the equivalence principle).
For atomic beam interference the influence of the homogeneous
gravitational acceleration has been shown by Kasevich and Chu
\cite{kasevich91} and Shimizu {\em et al.} \cite{shimizu92}. The
Sagnac effect for matter waves representing the influence of the
rotation of the reference frame has been measured for neutrons
and the rotating
earth by Werner {\em et al.} \cite{WSC}, for atomic beams on a
turntable by Riehle {\em et al.} \cite{riehle91}, and for
electrons
by Hasselbach and Nicklaus \cite{hasselbach88}.

Turning to the theoretical discussion we mention that the
measurability of the corresponding effects for atomic beam
interferometers has been discussed by Clauser \cite{clauser88}
and for the Ramsey interferometer by Bord\'e \cite{borde89}.
Mashhoon \cite{mashhoon88}, Silverman \cite{silv89b}, and
Audretsch and L\"ammerzahl \cite{aulae92} have pointed out that
in addition to the Sagnac effect a spin-rotation effect may
be measurable for neutrons and atoms, respectively. Audretsch
{\em et al.} \cite{aubllaem} have shown that Lorentz invariance
may be tested with atomic beam interferometry.

In this paper we discuss the next step in this context
and turn to the deepest
feature of gravity, to space-time curvature. Our main aim is to
demonstrate the measurability of the influence of curvature
on the dynamics of a quantum system under laboratory conditions,
a question also addressed by Anandan \cite{anandan84}
in the context of neutron interferometry.
Typical results could be a shift of energy levels or a
phase shift in an interference experiment. In the
laboratory the first
effect is hopelessly small whereas the curvature caused shift
of the optically induced Ramsey fringes in atomic interferometry
seems to be in the range of todays equipments.

This will be shown below. To prevent misunderstandings:

presently quantum systems
are not able to serve as probes which are capable to
discriminate better between alternative space-time theories of
gravity (PPN parameters) than astrophysical 'test particle`
systems like the binary pulsar do. The conceptual importance
of the interference experiment must rather be seen in the fact
that it will be possible to demonstrate that and how Einstein's
curvature tensor becomes effective within a quantum system
under laboratory conditions. This would be a very important
experimental investigation of the connection between quantum
theory, which is typically valid on microscopic scales, and
general relativity, which
describes well our world on very large scales. Or seen from
another point of view, this would be an experiment which
enlarges the range of validity of general relativity to
much smaller scales and especially to scales where the
classical description of matter breaks down.

Experimental quantum optics is a rapidly evolving field with
quick progress in the development of new methods of
measurement and the improvement of sensitivities. It therefore
seems to be most promising to examine also future possibilities
offered by this field of physics. Accordingly, it is the
second aim of this paper to give a short and transparent
derivation of the respective influences of the curvature
on different levels of approximation
starting from the Hamiltonian operator of the complete
theory of quantum mechanics in curved space-time based on the
general relativistic Dirac equation. The underlying physical
principles, the different approximation steps, and the
theoretical status of the experiment will thereby become clear.
This theoretical part is also intended to be a contribution to
{\em quantum optics in curved space-time}
which may be regarded as a special domain of
quantum optics under non-minkowskian conditions.

In addition to curvature we include below inertial influences
from the beginning as in Audretsch {\em et al.} \cite{AHL}.
The corresponding single effects
obtained on the final level of approximation are not new. This
unifying approach seems nevertheless to be justified
because other authors specialize on acceleration or rotation
or curvature only. An alternative approach based on the
quantization of the Hamiltonian of a point particle in a weak
gravitational field (deWitt \cite{dewitt66}, Papini \cite{papini67})
is less
general because it cannot handle spin effects.

The paper is organized as follows:\\
In Sec. 2 we derive the Dirac Hamiltonian with an
electromagnetic potential for stationary space-times and
rewrite it with reference to the Fermi coordinates attributed
to an accelerated, rotating observer. The influence of inertia
and curvature on the energy levels of a hydrogen atom is worked
out in Sec. 3. In Sec. 4 we go to the non-relativistic limit
and discuss the resulting correction terms in the Pauli
equation. For weak gravitational fields the curvature terms are
related to the Newtonian potential in Sec. 5. This serves in
Sec. 6 as a foundation of the discussion of the theoretical
relevance of the proposed experiments. In Sec. 7 we turn to the
experiments and work out the influence of acceleration
(as deviation from free fall),
rotation, and space-time curvature on the fringes of a Ramsey
interferometer. The resulting phase shifts are given for
different orientations of the interferometer. Finally, in Sec. 8
the orders of magnitude of the phase shifts are discussed for
existing experimental set-ups and modifications of the Ramsey
interferometers which should be technically
possible in the near future. For the gravitational field
we thereby refer to the field of the earth and alternatively to
the field of laboratory sized lead blocks.
Details of the calculations are
presented in three appendices.

Due to the twofold aim of the paper its sections may be

more or less interesting for physicists with different

research fields. We have taken this into account.
Those who are interested in the derivation
of the correction terms to the Pauli equation and not
in interferometry should focus on section 2 to 5 and the
appendices. Those who would like to see the experimental
implications may skip these sections and
start reading with section 6.

We use natural units ($\hbar = c= 1$) unless

otherwise stated.

\section{Dirac Hamiltonian and Fermi coordinates}
In a first step we derive the Dirac Hamiltonian in a stationary
curved space-time in a rigorous way and introduce Fermi
coordinates as a local coordinate system. We thereby generalize
the work of Parker \cite{parker80} to an observer in arbitrary
motion including rotation and acceleration.

A key assumption
of this approach is that hydrogen-like atoms can be modelled
by an electron in the given Coulomb field of the nucleus.
In another approach Fischbach, Freeman, and Cheng

\cite{fischbach81} have treated the hydrogen atom
as a two-body system in the gravitational field of the
earth and have found that there are differing correction terms
depending on whether one uses center of mass or center of energy
coordinates. Especially they got correction terms

to the Hamiltonian which do not vanish
even in the limit of very high proton to electron mass

ratio (Table I of Ref. \cite{fischbach81}).

For the purpose of interferometry this causes no difficulties
since these terms act on the internal degrees of freedom which
can be neglected in the phase shift (see below).

Since the calculation in the next two sections is close to that
of Parker we hold the presentation concise and refer to Ref.
\cite{parker80} and to our appendix A
for more details.

We rewrite the Dirac equation in an electromagnetic field with
four-vector potential $A_{\mu}$ in curved space-time in the
form

\beq i \partial_0 \psi = H \psi \label{dirac} \eeq
with
\beq H=-i (g^{00})^{-1} e^{\underline{\alpha}0}e^{\underline{

     \beta}i} \gamma_{\underline{\alpha}} \gamma_{\underline{

     \beta}} (\partial_i -\Gamma_i -iqA_i )+i \Gamma_0 -qA_0 -i

     (g^{00})^{-1}m e^{\underline{\alpha}0}\gamma_{\underline{

     \alpha}} \; . \label{diham} \eeq
Because of
\beq (\psi ,H \psi ) - (H \psi ,\psi ) = i \int d^3 x\,

     \psi^+ \gamma^{\underline{0}} \frac{\partial}{\partial

     \tau} (\sqrt{-g}\; e^0_{\; \underline{ \mu}}

     \gamma^{\underline{\mu}} )\psi \; ,\eeq
the operator $H$ is in general not hermitean with respect to
the conserved scalar product (\ref{sp}) given in the
appendix A. It can therefore be interpreted as the Hamiltonian

only in certain spacetime geometries, especially when the

metric is stationary. This reflects one of the numerous
conceptual problems which arise for quantum theory in curved
space-times (comp. Ref. \cite{fulling89}).
We assume in the following stationarity. $H$ of Eq.

(\ref{diham}) is then hermitean. It is then interpreted as the
Hamiltonian operator
representing the observable total energy of the massive
quantum object.

This can be seen more clearly if we
apply the method of L\"ammerzahl \cite{laemm83} and
refer to the energy-momentum
tensor $T_{\mu \nu}$ of the Dirac field:
\beq T_{\mu \nu} = - {1\over \sqrt{-g}} \frac{\delta {\cal L}}{
     \delta e_{\underline{\alpha}}^\nu} e_{\underline{\alpha}

     \mu} \eeq
(${\cal L}$ is the Lagrangian).
Let $\xi^\mu$ be the timelike Killing vector of the stationary
space-time and assume that the Maxwell field is also stationary.
Then the integration of $T_{\mu \nu}\xi^\mu$ over a spacelike

hypersurface
gives a conserved quantity which can be interpreted as the
energy of the Dirac field which includes the gravitational
potential energy. The Hamiltonian operator is defined by setting
its expectation value equal to the conserved energy
\beq E = \int_{\Sigma} T_{\mu \nu}\xi^\mu d\Sigma^\nu =:
     (\psi ,H \psi )\; . \eeq
The result is
\bea H \psi &=& \xi^\mu n_\mu \{ i \gamma^\nu n_\nu m \psi -

     i( \gamma^\nu \gamma^\sigma -g^{\nu \sigma}) n_{\sigma}
     \hat{D}_{\nu}\psi \} + i(\delta_\nu^{\; \mu} + n_\nu n^\mu
     ) \xi^\nu \hat{D}_\mu \psi \nn \\
     & &+ {i\over 4}(\gamma^\mu \gamma^\nu -
     g^{\mu \nu})\; (D_\mu \xi_\nu )\psi - q \Phi \psi

     \label{clausham} \eea
where $n_\mu$ is the normal vector of the hypersurface,
$\hat{D}_{\mu} = D_{\mu} - iq A_{\mu}$,
and $\Phi$ is defined by
\beq \partial_\mu \Phi = F_{\mu \nu} \xi^\nu \; .
     \label{phibed} \eeq

Choosing
the hypersurface $x^0 =$ constant, i.e., $n_\mu = (1,0,0,0)/
\sqrt{-g^{00}}$, and using the fact that

$\xi^\mu = (1,0,0,0)$ is a Killing vector if $\partial_0 g_{\mu
\nu}=0$ holds, we can easily prove the agreement of this

definition with the one of Eq. (\ref{dirac}), provided the
Maxwell field is stationary, $\partial_0 A^\mu =0$.
The stationarity condition on the Maxwell field prevents us
from running into troubles with gauge invariance as indicated by
Yang \cite{yang76}. Consider first the case of flat space-time.
The only term which is not gauge invariant is then the one
proportional to $A_0$ (in Eq. (\ref{clausham}) this corresponds
to the $\Phi$ term). The stationarity condition restricts the
gauge transformations to those of the form $A^\prime_{\mu} =
A_{\mu} + \partial_{\mu} \chi$ with $\chi = g(\vec{x}) + k x^0 $
for some constant $k$. A gauge transformation leads therefore
only to the addition of a constant $k$ to the Hamiltonian which
does not affect the eigenfunctions of it. In Eq.
(\ref{clausham}) the situation is even simpler since $\Phi$ is
determined by Eq. (\ref{phibed}) which is manifest gauge
invariant. The remaining freedom to add a constant to $\Phi$
corresponds to the effect of allowed gauge transformations in
flat space.

Turning to Fermi coordinates
we attribute to our quantum system an observer moving along
the worldline $P_0(\tau )$ with four-velocity

$u^\alpha (\tau )$. $\tau$ is its proper time along the
worldline. The experimental set-up (e.g. the interferometer) is
assumed to remain fixed with regard to a
{\em comoving orthonormal tetrad} $e^\alpha_{\; \underline{\mu}}$
representing a frame of reference
which is introduced as follows: $e^\alpha_{\underline{0}}$ is
identical with $u^\alpha$ and the orthonormal spatial triad
$e^\alpha_{\underline{i}}(\tau )\; ( i=1,2,3)$ rotates
together with the set-up with proper angular velocity $\vec{
\omega}$. Non-gravitational forces lead to a proper

three-acceleration $\vec{a}$ causing a deviation from the
trajectory of the free fall ($a^\alpha = \nabla_u u^\alpha$).
The gravitational field is represented by the Riemann curvature
tensor $R_{\mu \nu \rho \sigma}(\tau )$ along the worldline.

{\em Fermi coordinates} are the local coordinate system which is
designed to analyze experiments in the proper frame of reference

of an accelerated, rotating observer in curved space-time.
They are introduced as follows: the time coordinate on $P_0
(\tau )$ is the proper time of the
observer. The spatial coordinate lines are constructed by
sending out geodesics orthogonal to the observers world line
corotating with the spatial triad
$e^\alpha_{\underline{i}}(\tau )$. Each event near the
observers worldline is intersected by one spacelike geodesic
with tangent vector $n^\alpha$ orthogonal to $e^\alpha_{
\underline{0}}$ in the point $P_0(\tau )$ with observer time
$\tau$. Let $s$ be the distance of the event measured along
this geodesic. The Fermi coordinates of the event are then given
by $x^0 =\tau \; ,\; x^k = s n^\alpha e^{\underline{k}}_{
\alpha}$. At the observers worldline they represent a
rectangular grid attached to the experimental set-up which is
accelerated with $\vec{a}$ and rotating with $\vec{\omega}$
(comp. Ref. \cite{MTW}).

The metric in this

coordinate system is Minkowskian on the whole worldline
$P_0(\tau )$ and can

be calculated up to a given order in the spatial coordinates

$x^l$. To second order it is given by \cite{manasse63,ni78}
\bea g_{00} & = & -(1 + \vec{a} \cdot \vec{x})^2 + (
     \vec{\omega} \times \vec{x})^2 - R_{0l0m}x^l x^m +

     O((x^l)^3)\nn \\ g_{0i} & = & \varepsilon_{ijk} \omega^j

     x^k - {\st \frac{2}{3}} R_{0lim} x^l x^m  \nn +

     O((x^l)^3)\\ g_{ij} & = & \delta_{ij} -{\st \frac{1}{3}}

     R_{iljm} x^l x^m + O((x^l)^3) \label{metrik} \eea
whereby $\vec{a}$, $\vec{\omega}$, and the curvature tensor
$R_{\mu \nu \rho \sigma}$ are taken on the
observers worldline. They
may depend on $\tau$. The use of Eq. (\ref{metrik})
represents an approximation which is good as long as the
characteristic dimension $s$ of the quantum system is small
compared to the characteristic lengths attributed to inertia
and curvature:
\beq s \ll \mbox{min } \left \{ \frac{1}{|\vec{a}|}\; ,\;
     \frac{1}{|\vec{\omega}|}\; ,\; \frac{1}{|R_{\mu \nu \rho
     \sigma}|^{1/2}}\; ,\; \frac{|R_{\mu \nu \rho \sigma}|}{|
     \partial_{\lambda} R_{\mu \nu \rho \sigma}|} \right \}

     \label{naehbed} \eeq
The expressions for the curvature given in appendix C show
that this is very well fulfilled
for quantum systems up to extremely strong curvature.

We rewrite Eq. (\ref{diham}) with reference to Fermi coordinates
using the results of appendix A and obtain for the Hamiltonian
operator
\bea H & =& -im \{ [1+ \vec{a}\cdot \vec{x} +
     {\st \frac{1}{2}}R_{0l0m} x^l x^m ]\gamma_{\underline{0}}

     +{\st \frac{1}{6}} R_{0lim} x^l x^m \gamma_{\underline{i}}

     \} - \vec{\omega} \cdot \vec{J}_0 + q \vec{x}\cdot
     (\vec{A} \times \vec{\omega})
     -q A_0 \nn \\ & &
      + \{ \alpha_{\underline{i}}
     + \vec{a} \cdot \vec{x} \alpha_{\underline{i}}
     + {\st \frac{1}{2}} [ R_{0lim}+R_{0l0m}\alpha_{
     \underline{i}} +{\st \frac{1}{3}}R_{iljm}\alpha_{
     \underline{j}} +{\st \frac{1}{3}}R_{0ljm}
     \gamma_{\underline{j}}\gamma_{\underline{i}}] x^l x^m \}
     (-i \partial_i -q A_i) \nn \\ & &

     -{\st \frac{i}{2}}a_i \alpha_{\underline{i}}
     +{\st \frac{i}{4}}\gamma_{\underline{i}}\gamma_{

     \underline{j}}

     R_{0imj} x^m +{\st \frac{i}{4}}\alpha_{\underline{j
     }}[R_{jm}-R_{0j0m}]x^m
\label{Hamilton} \eea
where $\vec{J}_0 = -i (\vec{x}\times \vec{\nabla})

+\vec{\Sigma}/2 $ is the total angular momentum in absence of an

electromagnetic field. We have used $R_{mijk}

\varepsilon_{ijk} =0$ and
\beq \gamma_{\underline{i}}\gamma_{\underline{j}}

     \gamma_{\underline{k}} = \delta_{ij}\gamma_{\underline{k}}

     -\delta_{ik}\gamma_{\underline{j}} +
     \delta_{jk}\gamma_{\underline{i}} +i \varepsilon_{ijk}

     \gamma_{\underline{5}}\gamma_{\underline{0}} \label{3gamma}

     \eeq
\section{Energy shift of the hydrogen atom}
Once the complete Hamiltonian is given one can use time

independent perturbation theory for degenerate states to

calculate the influence of $\vec{a}$, $\vec{\omega}$, and
$R_{\mu \nu \rho \sigma}$ on the energy levels of the
hydrogen atom. The knowledge of these energy corrections is
important also in connection with Ramsey interferometry since we
have
to answer the question whether the corresponding
modification of the internal
degrees of freedom have to be taken into account or not. They
are given to

lowest order by the eigenvalues of the matrix $<\alpha
| H_{pert}| \beta >$, where $H_{pert}$ is the difference between

Hamiltonian in Fermi coordinates and the Hamiltonian $H_0$ in
Minkowski space
\beq H_0 = -i \alpha_{\underline{i}} \partial_i +\frac{qZe}{r}

     -im \gamma_{\underline{0}} \eeq
and $|\alpha>$ and $|\beta>$ are two degenerate states of a
hydrogen-like atom written as Dirac spinors. We assume that the
multiplicity of the unperturbed energy level is two.

To calculate the shift of the energy for the ground state

of the hydrogen atom we have to insert the field of a point

charge resting at the origin of the Fermi coordinate system
into the Hamiltonian. In flat space this is the
ordinary Coulomb field, but the presence of curvature and the
non-inertial motion of the atom lead to corrections to the
Coulomb potential which have to be included in the energy
calculation. In appendix B we give a derivation of these
correction terms.

Generalizing the calculation of Parker \cite{parker80} we find
for the energy shift of the ground state
\beq \bigtriangleup E = \frac{2 \gamma +1}{6 m}\left \{

     \vec{\omega}^2 -\vec{a}^2
     + \frac{1}{4}R +R_{00}\left ( \frac{3}{2}+ \frac{\gamma

     (\gamma +1)}{2
     \zeta^2}\right ) \right \} \pm  \frac{1}{2}| \vec{\omega} | \eeq
where $\zeta = -qZe$ and $\gamma = (1-\zeta^2 )^{1/2}$.

Most of these energy shifts are far outside the measuring range
of modern experiments. If one assumes Einstein's field equations
$R$ and $R_{00}$ vanish in a vacuum where the
experiments should be performed. The terms quadratic in the

rotation and acceleration give the contributions $8.4 \cdot

10^{-37} eV \cdot (\omega / 1 Hz)^2 $ and
$9 \cdot 10^{-52}eV \cdot (a/g)^2$ where $g=9.81 m s^{-2}$.
The last term, caused by rotation, has a
magnitude of $6.6 \cdot 10^{-16} eV \cdot \omega /1 Hz$. This
may be big enough to be detected via an induced optical activity
in atoms \cite{silv89}.

To return to interferometry: supposing one can
manage that the time of flight of the atoms is very long,

say one second ($1.5 \cdot 10^{15} eV^{-1}$ in natural units),
we can see that the energy shift of the ground state results
only in a very tiny phase shift $\Delta \phi$ by setting
approximately

$\Delta \phi = \Delta E \cdot t$. Again, only the last

rotational term may cause measurable effects. For Ramsey
interferometry we may therefore refer to the unperturbed energy
levels of the atoms.
\section{Non-relativistic limit}
In general the phase shift of an interferometric
pattern caused by some external force increases with decreasing
velocity of the particle beam. It is therefore adequate to
derive a non-relativistic approximation of the Hamiltonian
which leads us at last to the modified Pauli equation. In a
systematic way this is best done by use of the {\em
Foldy-Wouthuysen transformation} FWT  (see, e.g., Ref.
\cite{ItZu}). The idea of this transformation is

to construct a unitary transformation $\psi = \exp (-iS)

\psi^\prime $ with $(\phi ,S \psi ) = (S\phi ,\psi)$ such that

in the Operator $H^\prime$ acting on $ \psi^{\prime}$,
\beq H^\prime  \approx H + i

     [S,H] - \frac{1}{2} [S,[S,H]] -\dot{S} -\frac{i}{2}

     [S,\dot{S}] + \cdots \; , \label{FW} \eeq
the odd operators, which couple the small components of the

spinor to the large ones, are relativistically suppressed,

i.e. are of higher order in $1/m$. Because of this intended
separation of the large and small spinor components the
calculations are usually done in the standard representation
of the Clifford algebra in which, with
our conventions for the sign of the metric, $\gamma_{\underline{
0}} = i $ diag$(1,1,-1,-1)$ holds.
It should be stressed that $S$

must be hermitean with respect to the scalar product
(\ref{sp}) in curved

spacetime in order to preserve the respective hermiticity of

transformed operators. In Fermi coordinates the scalar product
is given by Eq. (\ref{skalprod}).

We now proceed with the derivation of the FWT in

Fermi coordinates in constructing the operator $S$.
In Minkowski space it is chosen to be $\gamma_{\underline{j}}
(-i \partial_j -qA_j) /2m$ in order to suppress the

$\alpha_{\underline{i}} (-i \partial_i -qA_i)$ term. Its hermitean
generalization
in Fermi coordinates is
\beq S= \frac{1}{2m}[\gamma_{\underline{j}} (-i \partial_j
     -q A_j) + {\st

     \frac{1}{6}} R_{0lim} x^l x^m \; \gamma_{\underline{0}}
     \gamma_{\underline{i}} \gamma_{\underline{j}}
     (-i \partial_j -q A_j) +{\st

     \frac{i}{6}}R_{ijim}x^m \; \gamma_{\underline{j}}] \eeq

and leads, to lowest order in each perturbation, to
\bea H^\prime &=& -im(1 +\vec{a}\cdot \vec{x} +{\st \frac{1}{2}}

     R_{0l0m}x^l x^m )\gamma_{\underline{0}} -{\st \frac{i}{6}}m

     R_{0lim}x^l x^m \; \gamma_{\underline{i}}

     \nn \\ & & -{\st \frac{i}{2m}} \gamma_{\underline{0}}

     (-i \partial_i -qA_i)\, (-i \partial_i -q A_i)
     -q A_0 \nn \\ & &

     -\vec{\omega} \cdot \vec{J}_0 + q \vec{x}\cdot (\vec{A}
     \times \vec{\omega}) +  i{\st \frac{q}{2m}}
     \gamma_{\underline{0}} \vec{\Sigma} \cdot \mbox{rot}
     \vec{A} +{\st \frac{q}{2m}} \gamma_{\underline{j}}

     (\dot{A}_j -\partial_j A_0 )

\label{zwstufe} \eea
It is remarkable that, though $S$ does not depend on the

acceleration, all odd terms containing $\vec{a}$ are removed to

this order.

The Hamiltonian $H^\prime$ still contains two odd terms.
The last one ($ \propto q \gamma_{\underline{j}}$) can be
treated as in the textbooks
\cite{ItZu}. We will omit this. The first one
($ \propto R_{0lim}$) needs particular attention. It
cannot be removed by a second FWT, because only the combination
\beq -im (\gamma_{\underline{0}} + {\st \frac{1}{6}} R_{0lim}x^l

     x^m \gamma_{\underline{i}}) \eeq
in $H^\prime$ is hermitean. If a hermitean operator $S$
suppressing the $R_{0lim}$ term would exist it would produce a
non-hermitean Hamiltonian in contradiction to the scheme.

To solve the problem in a consistent way we propose the
following approach: Although we refer
to the quasi-cartesian Fermi coordinates we still have
to make use of the scalar product (\ref{skalprod}) which does
not agree with the non-relativistic scalar product in
Cartesian  coordinates. To make
the correspondence complete we look for an operator
$O$ which transforms $\psi^\prime$ to $\tilde{\psi}=O^{-1}
\psi^\prime$ in such a way that the scalar product is

changed to the one in flat space:
\beq (\phi^\prime ,\psi^\prime ) = (O

     \tilde{\phi},O\tilde{\psi}) = (\tilde{\phi},\tilde{\psi})_0

     \equiv \int d^3 x\; \tilde{\phi}^+ \; \tilde{\psi}

     \label{sbeding} \eeq
Given this condition the conserved scalar product remains
conserved after the transformation.
One can do this transformation formally for arbitrary

coordinates and the general scalar product (\ref{sp})  which
may be written as
\beq (\phi^\prime ,\psi^\prime ) = (\phi^\prime ,T \psi^\prime
     )_0 \; , \quad T:= -\sqrt{-g}\, e^0_{\; \underline{\mu}}
     \gamma^{\underline{0}}\gamma^{\underline{\mu}} \eeq
The condition (\ref{sbeding}) for $O$ then takes the form
\beq (\phi^\prime ,\psi^\prime ) = (O \tilde{\phi} , TO \tilde{
     \psi})_0 \stackrel{!}{=} (\tilde{\phi},\tilde{\psi})_0 \eeq
where the last equality is the demand. This leads immediately to
\beq O^{*T} T O = 1\; ,\quad O^{*T} T = O^{-1} \eeq

Accompanying the transformation of states with the corresponding
transformation $A^\prime \rightarrow \tilde{A}=O^{-1}A^\prime
O$ of operators we obtain
\beq (\phi^\prime , A^\prime \psi^\prime ) = (\tilde{\phi},
     O^{*T} T  A^\prime O \tilde{\psi})_0 =

     (\tilde{\phi},\tilde{A} \tilde{\psi})_0 \eeq
Similarly one can show that
\beq (A^\prime \phi^\prime , \psi ^\prime ) = (\tilde{A}

     \tilde{\phi} , \tilde{\psi})_0 \eeq
Using these two equations it is not hard to see that if any

operator $A^\prime $ is hermitean with respect to the
curved scalar product, $(\phi^\prime , A^\prime \psi^\prime )
= (A^\prime \phi^\prime , \psi^\prime )$, then so is

$\tilde{A}$ with respect to $(,)_0$. In the same sense the
unitarity of an operator is conserved under this transformation.
Because of its properties the product changing transformation
will be called {\em quasi-unitary}.
Of great importance for the consistency of the FWT is the
fact that the ordering of the two transformation $\exp (iS)$ and
$O$ plays no role since
\beq O^{-1} e^{iS} A^\prime e^{-iS} O = O^{-1} e^{iS} O

     \tilde{A} O^{-1} e^{-iS} O = e^{i \tilde{S}} \tilde{A}
     e^{-i \tilde{S}} \eeq
If we perform the quasi-unitary transformation first we simply
have to take the transformed operator $\tilde{S}$ to do the
FWT. This sequence is in fact more convenient since the

hermiticity of $\tilde{S}$ with respect to $(,)_0$ is easier

checked than the one of $S$ with respect to the curved scalar
product.

For the general scalar product (\ref{sp}) the operator $O$
takes the form
\beq O = P + Q e^{0 \underline{i}} \alpha_{\underline{i}} \eeq
where
\beq P= \frac{\dst 1}{\dst \sqrt{2 g_\Sigma}} [ \sqrt{g_\Sigma}

   -\sqrt{-g}\; e^{0 \underline{0}} ]^{1/2} \eeq
and
\beq Q = \frac{\dst -\sqrt{-g}}{\dst \sqrt{2

     g_\Sigma}}[\sqrt{g_\Sigma}-
     \sqrt{-g}\; e^{0 \underline{0}} ]^{-1/2} \eeq
$g_\Sigma$ is the determinant of the spatial part of the metric.
With respect to Fermi coordinates and the related approximation,
the operator $O$ becomes
\beq O = 1 + \frac{1}{12} R_{ilim}x^l x^m + \frac{1}{12}

     R_{0lim}x^l x^m \; \alpha_{\underline{i}}+ O((x^l)^3) \; .

     \eeq

Application to $H^\prime$ of Eq. (\ref{zwstufe}) removes the
curvature induced odd term so that we finally find for the

Hamiltonian in the non-relativistic limit the result
\bea \tilde{H} &=& -im(1+\vec{a}\cdot \vec{x} +	{\st

     \frac{1}{2}}R_{0l0m}x^l x^m )\gamma_{\underline{0}} -{\st

     \frac{i}{2m}} \gamma_{\underline{0}}

     (-i \partial_i -qA_i)\, (-i \partial_i -qA_i)
     - qA_0 \nn \\ & & -

     \vec{\omega}\cdot \vec{J}_0  +q \vec{\omega}\cdot

     (\vec{x} \times \vec{A}) +i{\st

     \frac{q}{2m}} \gamma_{\underline{0}} \vec{\Sigma}\cdot

     \mbox{rot}\vec{A} + {\st \frac{q}{2m}}

     \gamma_{\underline{j}} (\dot{A}_j -\partial_j A_0)
\label{endeth} \eea
where $\vec{J}_0 = -i (\vec{x}\times \vec{\nabla})

+\vec{\Sigma}/2 $ is the total angular momentum in absence of an

electromagnetic field. The correction terms in the Pauli

equation are thus

\beq m \vec{a}\cdot \vec{x} \label{pertacc} \eeq

for acceleration,

\beq -\vec{\omega}\cdot \vec{J}_0 + q \vec{x} \cdot (\vec{A}

     \times \vec{\omega}) \label{pertom} \eeq

for rotation, and

\beq \frac{m}{2} R_{0l0m}x^l x^m  \label{pertcurv} \eeq

for curvature. The term proportional to the rotation and the
electromagnetic field can be understood to arise from the
minimal coupling of the spinor and the four potential
so that no problem with gauge invariance occurs.
Each term is hermitean with respect to the

Schr\"odinger type scalar product $(,)_0$ of Eq. (\ref{sbeding})
in Cartesian coordinates even if the term is time dependent.
Due to
the approximation (\ref{metrik}) of the metric, which is correct
only to second order in the spatial coordinates,
we find no mixing

between $\vec{a},\vec{\omega}$ and $R_{\mu \nu \rho \sigma}$.
Only $\vec{a}$ and $\vec{\omega}$ are mixed in the $g^{0i}$
components of the metric, and this leads to a corresponding
term in the vector potential (\ref{mixing}).

A higher expansion of the metric \cite{li79} leads also to a
mixing between $\vec{a}$, $\vec{\omega}$ and $R_{\mu \nu \rho
\sigma}$ and would presumably produce corresponding effects
in the vector potential and in the Schr\"odinger equation.

Our results for acceleration and rotation are in agreement
with previous calculations \cite{overh74,schmutzer73,hehl90}
and have an obvious physical interpretation (comp. Ref.
\cite{aulae92}).

The curvature term confirms Parker's \cite{parker80}
approximative approach to the

hydrogen atom and

coincides with another derivation which, instead of starting
from a covariant quantum theory, takes
a classical point particle in a weak gravitational field
as starting point \cite{dewitt66,papini67,brillet83}.
\section{Weak gravitational fields}
For later use we give a physical interpretation of the
curvature term in
relating it to the Newtonian potential in the
limit of weak gravitational fields (comp., e.g., chapter 18
of \cite{MTW}). To do so we turn in a last step to
the linearized approximation of
general relativity which is fulfilled in the solar system or for

gravitational waves. In this case the metric is written as
\beq g_{\mu \nu} = \eta_{\mu \nu} + h_{\mu \nu} \; , \quad
     |h_{\mu \nu}| \ll 1 \eeq
Defining
\beq \bar{h}_{\mu \nu} := h_{\mu \nu} - {1\over 2} \eta_{\mu
     \nu} h^{\rho}_{\; \rho} \eeq
and choosing harmonic coordinates so that $\bar{h}^{\mu \alpha}
_{\; \; \; ,\alpha} =0$ the linearized field equations become
\beq -\partial_0 \partial_0 \bar{h}_{\mu \nu} +\sum_{i=1}^3

     \partial_i \partial_i \bar{h}_{\mu \nu} = - 2 \kappa T_{\mu

     \nu}\eeq
where $T_{\mu \nu}$ is the energy-momentum tensor of the matter

producing the gravitational field, and
the constant $\kappa$ is defined to be $\kappa = 8 \pi G$

whereby $G$ is Newtons constant. Static nearly Newtonian
sources obey $T_{00} \gg |T_{0j}|\; , \; T_{00} \gg |T_{jk}|$
with $T_{00} = \rho (\vec{x})$.
In this case the solution of the field equation is $\bar{h}_{0j}
= \bar{h}_{jk}=0$ and
\beq h_{00}(\vec{x})=h_{i(i)}(\vec{x})= \frac{1}{2}

     \bar{h}_{00}(\vec{x}) = 2

     G \int \frac{\rho (\vec{x}^\prime )}{|\vec{x}

     -\vec{x}^\prime |} \, d^3 x^\prime \label{naehe} \eeq
which corresponds exactly to the {\em Newtonian potential}:
$\phi (\vec{x})= -h_{00}/2$.

Let us consider an observer resting in

this space, $\dot{x}^i =0$. Without loss of generality we can
take $\vec{x} = 0$. From
Eq. (\ref{ngeod})  his acceleration is found to be
\beq a^0 = \ddot{x}^0 =0\; ,\; a^i = \partial_i \phi (0)\,

     (\dot{x}^0)^2 =\partial_i \phi (0)\eeq
We see in accordance with the interpretation given in Sec. 2
that $m \vec{a}$ is exactly the force needed to

resist the Newtonian gravitational force, i.e., the negative of
it. Note that, due to

$\dot{x}^\mu \dot{x}_\mu = -1$, the equation
\beq x^0 (\tau ) = \tau (1- \phi (0))
     \label{eigenzeit} \eeq
holds. The components of the curvature tensor take the values
\beq R_{\mu \nu \alpha \beta} = {1\over 2} [ \partial_\nu

     \partial_{\alpha} h_{\mu \beta} + \partial_{\mu}
     \partial_{\beta} h_{\nu \alpha} - \partial_{\mu}
     \partial_{\alpha} h_{\nu \beta} - \partial_{\nu}
     \partial_{\beta} h_{\mu \alpha} ] \eeq
so that
\beq R_{0l0m} = \partial_l \partial_m \phi (\vec{x})
     \label{krupot} \eeq
These curvature components are gauge invariant to linear terms
in $h_{\mu \nu}$  so that the components in the Fermi coordinate
system agree with those of (\ref{krupot}).
This illuminates the physical meaning of the curvature term
(\ref{pertcurv}).

One may look at these results also from a different point of
view starting with the Newtonian potential.

If we expand $m \phi (\vec{x}) $ around $\vec{x} =0$

we get
\bea m \phi (\vec{x}) &=& m \phi (0) +
     m x^l  \partial_l \phi (0) + {m\over 2} \partial_l
     \partial_m \phi (0)\, x^l x^m + O((x^l)^3 ) \nn \\
     & = & m \phi (0) + m a^l x^l  +{m\over 2} R_{0l0m}(0)\,

     x^l x^m + O((x^l)^3 ) \label{taylor} \eea
so that, in this case, the acceleration term and the

curvature term in the Hamiltonian are simply the first terms of
the Taylor expansion of the
Newtonian potential difference. The potential $\phi (0)$ on the
worldline of the observer is absorbed into the proper time
$\tau$ via (\ref{eigenzeit}) and does therefore not appear
if the Hamiltonian is written in Fermi coordinates.
\section{Curvature and Newtonian gravity}
The basic tool used above to mesh quantum mechanics with
gravity was Einstein's equivalence principle which demands that
in a local inertial frame (local Lorentz frame) all laws of
physics must take their special-relativistic form. This leads
to the general-relativistic Dirac equation which fixes the
dynamics. Because operators and states are attributed to
hypersurfaces of equal 'time` additional considerations were
necessary. Following a succession of approximations we obtained
the Hamiltonian (\ref{endeth})
for non-inertial reference frames which includes the space-time
curvature components
$R_{0l0m}$. For weak static gravitational fields
$R_{0l0m}$ may be related according to (\ref{krupot}) to an
inhomogeneous Newtonian gravitational potential $\phi
(\vec{x})$.

The three approximation steps were imposed on us by the
characteristic physical scales of quantum systems in
non-inertial frames in the solar system (in particular on and
near to the earth):
\\(i) The extension of the quantum system is small compared to
the characteristic lengths of curvature, acceleration, and
rotation (comp. Eq. (\ref{naehbed}). A treatment in Fermi
coordinates is therefore physically well justified.\\
(ii) In atomic beam spectroscopy atoms are very slow. The
Foldy-Wouthuysen transformation provides also in curved
space-time a systematic approach
to the non-relativistic limit.\\
(iii) The interferometer experiments will be done in a
laboratory on earth. Because of the precision which can be
obtained today we will not be able to test the respective
post-newtonian terms which would correct relation
(\ref{krupot}). With
the exact expressions for $R_{\mu \nu \rho \sigma}$ provided
in appendix C it would be easy to work out these corrections
for EinsteinBs theory.

In going to the correspondence limit of weak gravity and low
velocities of the quantum objects, have we lost the concept
of space-time curvature in this way? Equation (\ref{taylor})
demonstrates that this
is not the case. Space-time curvature manifests itself as
inhomogeneous gravitation and vice versa. The trajectories of freely
falling test particles

are geodesics of space-time. Curvature shows up in the deviation
of two nearby geodesics. This effect is physically equivalent
to the relative acceleration of the test particles. The
respective tide-producing gravitational forces are expressed in
terms of the curvature tensor $R_{\mu \nu \rho \sigma}$.
Space-time flatness would be equivalent to the absence of
tidal gravitational forces. But there is relative acceleration
of freely falling test particles in the limit of weak
gravitational fields. Accordingly, there is Einstein curvature
also in this case showing the same characteristic influence
on matter as in strong gravitational fields. Its particular
influence on quantum objects has been described above.

It has been pointed out by Misner {\em et al.} (p.~305 of
Ref.~\cite{MTW}) that the
deepest features of Newtonian gravity are (1) the equivalence
principle and (2) space-time curvature. They manifest themselves
in the second and third term of the Taylor expansion of the
potential $\phi (\vec{x})$ as can be seen in (\ref{taylor}).

It must be stressed that this Taylor expansion is tied to
the Newtonian picture of gravity. Seen in this way curvature
is simply the next Taylor coefficient of the potential.
In general relativity, however, acceleration and curvature
have a completely different origin. Acceleration is something
which is bound to the motion of the observer (see Eq.
(\ref{ngeod})) and is produced by any {\em non gravitational}
forces. Accelerational effects arise if we describe nature
in the natural frame of reference or the Fermi coordinates
of the accelerated observer. In this frame freely falling,
forceless objects (like the apple) seem clearly to be
accelerated because their speed relative to the accelerated
observer is changing. Curvature, on the other hand, is {\em the}
quantity which describes the deviation of space-time from a
flat manifold. It has nothing to do with the observer or his
motion in space. Since general relativity describes the world
better than Newtonian gravity on large scales this point of
view is more appropriate. It follows that the measurement of
the influence of tidal forces on quantum systems would be a
qualitatively new contribution to the understanding of the
microscopic world. The influence of the acceleration
of all observers resting on the earth has been
tested in quantum mechanics whereas the influence of
space-time itself via  curvature is only established on large
scales, in a classical region, where the problems of the
description of quantum matter in a curved space are absent.

The validity of the equivalence principle for quantum systems
has already been tested, see section 1. To demonstrate that
not only classical test particles but quantum systems, too,
react in a measurable way on tidal forces and

therefore on space-time curvature is the aim of this paper.

The gravitational field available in a laboratory is the field
of the earth. There are two possible types

of  motion of the interferometer. It may be at rest at the

surface of the rotating earth or freely falling either on a path
towards the earth or fixed to an orbiting satellite. The
difference between the first and the last two cases is a
Lorentz boost and possibly a rotation. To obtain the order of
magnitude of the effects in question
it is therefore sufficient to consider different orientations
of the interferometer in a laboratory
on earth because the relative velocity of a flying laboratory
is so small that the respective relativistic corrections
caused by the boost will
not show up. We now turn to the description of the
interferometer.
\section{Phase shift of the Ramsey fringes}
With the results (\ref{pertacc}) to (\ref{pertcurv}) of section
5 we have the ability to examine the influence of acceleration,

rotation, and space-time curvature on non-relativistic
experimental set-ups. With regard to a demonstration of the
influence of curvature, the {\em Ramsey
atom beam spectrometer} seems to be
well suited. Most experiments done with this set-up are
concerned
with high resolution spectroscopy. But recently Bord\'e
\cite{borde89} has pointed out that it can also be used as an
atom interferometer. It is this aspect which we are
interested in.

To describe the apparatus we fix the comoving observer tetrad
to the interferometer so that the respective Fermi coordinates
can be interpreted
approximately as the ordinary Cartesian coordinate system
$(x^1, x^2, x^3)$ with range over the apparatus. The set-up
consists of an atomic beam which moves initially in the
$x^1$-direction and
four traveling laser waves parallel or anti parallel to the
$x^3$ direction. The laser waves are tuned to be nearly in
resonance with a particular transition between two states of the
atoms. The first two laser beams are copropagating in the
$x^3$-direction, the third and the fourth beam copropagate
in the $-x^3$-direction (see Fig. 1).

The time of flight $T$ between the
two lasers of a copropagating pair is the same for each pair.
Between the two pairs the atoms move for a time $T^\prime$.
While passing
the lasers, the atoms absorb or reemit photons. By the
corresponding recoil the atomic wavefunction is coherently
splited and recombined so that the two pairs of interfering
atomic beams of Fig. 1 are obtained. The outgoing wave functions
$b^{(+1)}$ and $b^{(-1)}$ correspond to an excited part of the
wave function. The index denotes the pair. The respective
population can be read out by detection of its fluorescence
radiation. It
oscillates with the laser detuning. The resulting oscillations
are called {\em Ramsey fringes}. With regard to their
resolution note that the excited atoms (dashed lines in Fig. 1).
may decay according to their lifetime $\tau_l$. For further details
of the experiment
and its theory see Bord\'e {\em et al.} \cite{borde84}.

In a previous paper \cite{maau92} we have calculated the
first order contribution to  the shift
$\Delta \varphi$ of the Ramsey fringes due to a fairly general
perturbation
\begin{equation} H^{(1)} = \hat{H}(\vec{p}) \; (x_1)^{N_1}(x_2)^{N_2}

     (x_3)^{N_3} \label{stoer} \end{equation}
proportional to powers of the coordinates with $N_i$

being integer.
$\hat{H}(\vec{p})$ contains an arbitrary dependence on the
center of mass momentum operator $\vec{p}$.
In a not yet published paper \cite{kpmunpub} higher orders
of non Minkowskian influences have been taken into account.
It is important to note that these calculations
are not made within the usual
WKB approach but is based on time dependent

first order perturbation theory of
the Schr\"odinger equation. Accordingly, one does not deal

with classical paths but follows a unitary time evolution of
quantum states. Any bending of a beam is automatically taken
care of. It does not appear
explicitly in the calculation. In addition, the time evolution
implies that  the specifications of the experiment refer to
the times of the
interaction between atoms and light and not to spatial
locations of the atoms.

The horizontal axis in Fig. 1 describes this time
evolution. The vertical axis is in principle only a schematic
drawing of the time evolution of the atoms. For sufficiently
localized wavepackets, however, it can approximately
be interpreted as
the relative distance between two partial beams. Since atoms
get simply a kick, i.e., some definite transfer of momentum
by the lasers, this relative distance grows linearly
in time even if a homogeneous gravitational field (acceleration)
is present.

The resulting shifts
$\Delta \varphi^{(-1)}$ and $\Delta \varphi^{(+1)}$
corresponding to the two interferometer geometries of Fig. 1
are given in Ref. \cite{kpmunpub} and
equations (28) and (31) of Ref. \cite{maau92}
(note the change in the notation). Specializing $H^{(1)}$ of
Eq. (\ref{stoer}) to our correction terms (\ref{pertacc}) to
(\ref{pertcurv}) we obtain for the
first order {\em phase shift}
\beq \Delta \varphi^{(\pm 1)} = \Delta \varphi^{(\pm 1)}_a +
     \Delta \varphi_{\omega}^{(\pm 1)} + \Delta

     \varphi_R \label{gesshift} \eeq
The influence of the acceleration $\vec{a}$ is given by
\beq \Delta \varphi^{(\pm 1)}_a \equiv
     \Delta \varphi_a = -k a_3\;  T (T+T^{\prime})     \label{acc}
\eeq
A rotation $\vec{\omega}$ leads to
\beq \Delta \varphi_{\omega}^{(\pm 1)}\equiv \Delta
     \varphi_{\omega} = {2\over m}
     \omega_2 p_1 k\; T (T+T^{\prime}) \label{rot} \eeq
Both shifts agree for the two interferometer geometries
(comp. also Ref. \cite{borde92}).
The phase shift $\Delta \varphi_R$ which is caused by space-time
curvature is the sum $\Delta \varphi^{(\pm1)}_{0103}
+\Delta \varphi^{(\pm 1)}_{0303}+ \Delta \varphi_{aR}$ of the
terms
\beq \Delta \varphi^{(\pm1)}_{0103} \equiv
     \Delta \varphi_{0103} = -{c^2 \over 2m} R_{0103}\; k
     p_1 T(2T^2 +3 T T^{\prime}+ (T^{\prime})^2)

     \label{0103} \eeq
\beq \Delta \varphi^{(-1)}_{0303} = - {\hbar c^2 \over 2m}
     R_{0303}\; k^2 (\frac{2}{3}T^3 + T^2 T^{\prime})

     \label{0303m}\eeq
\beq \Delta \varphi^{(+1)}_{0303}=- {\hbar c^2 \over 2m}
     R_{0303}\; k^2 T (4T^2 + 6T T^\prime + 3 (T^\prime)^2)
     \label{0303p} \eeq
\beq  \Delta \varphi_{aR} = {c^2\over 12} R_{030m} k a_m T
     (T+T^\prime)(7T^2+7TT^\prime+2T^{\prime 2}) \; .
     \label{aR} \eeq
For convenience we have reintroduced $\hbar$ and $c$ and have
set $k = 2\pi / \lambda$ where $\lambda$ is the wavelength of
the lasers. $v_1$ is the mean velocity of the atoms incident in
the $x^1$-direction and $p_1$ the corresponding

initial momentum.

The calculation
of the phase shift done in Ref. \cite{maau92} includes only
the derivation of Eqs. (\ref{0103}) to (\ref{0303p}). The mixed term
(\ref{aR}) which includes also the acceleration
must be calculated in a different manner.
Details will be given elsewhere \cite{kpmunpub}.
Because this expression cannot be found in an already
published paper we give some heuristic
arguments which may clarify its structure.
Note that the phase shift is a pure number. If
curvature is (to first order) included $\Delta \varphi_{aR}$
must be
proportional to $R_{0l0m} v_l w_m$ where $v$ and $w$ are certain
vectors. Furthermore, $\Delta \varphi_{aR}$
is an interference effect
and must therefore include something which indicates that the
atomic beam was splited and recombined. The only quantity
which can do this is the wave vector $k_m$ of the lasers.
Hence, $\Delta \varphi \propto R_{0l0m} k_l $. The rest of the
argument is based on the comparison with the curvature shifts
(\ref{0103}) to (\ref{0303p}). If the atoms were classical
balls there momentum in an accelerated frame of reference
would change according to $\vec{p} \rightarrow \vec{p}+M \vec{a}
T$. If we do this replacement for $p_1 \vec{e}_1$ or $\hbar k
\vec{e}_3$ in these equations we get an expression of the form
(\ref{aR}). Only the exact form of the fourth order
polynomial in the flight times $T, T^\prime$ differs.

In Ref. \cite{kpmunpub} a non-perturbative approach is
performed. It leads to the interesting result that all
acceleration dependent phase shifts are linear in $\vec{a}$.
No powers of $\vec{a}$ can appear if only acceleration,
rotation, and curvature are considered.
This is one point why it may be of advantage to use atoms
instead of neutrons. The work of Anandan \cite{anandan84}
demonstrates clearly that for neutrons there are several
contributions to the phase shift from different orders of the
acceleration alone so
that it may become more difficult to distinguish between
acceleration and curvature effects. Note also that the approach
of Anandan uses the WKB approximation and is therefore
completely different from our calculations which are based on a
unitary time evolution and make no reference to classical
paths.
\section{Estimation of the magnitude of the
experimental effects}
In the following we want to discuss in detail the measurability
of the influence of space-time curvature on the phase shift.
We thereby take as a basis the specifications of two already
existing experimental set-ups described by Riehle {\em et al.}
\cite{riehle91} and Sterr {\em et al.} \cite{ertmer92}. In Ref.
\cite{riehle91} the intercombination transition $^3 P_1
\rightarrow \, ^1 S_0$ of $^{40}$Ca has been used. In Ref.
\cite{ertmer92} the laser waves are resonant with the
intercombination transition $^3 P_1 \rightarrow \, ^1 S_0$ of
$^{24}$Mg. In the first and third row of table I the lifetime
$\tau_l$ of the excited metastable $^3 P_1$, the wavelength
$\lambda$ of the transition,
the mass $m$, and the initial velocity $v_1$ of the atoms

in the $x^1$-direction as well as
the times of flight $T,T^\prime$ between the laser beams in the
respective apparatus are given. The question mark in the
third row means that we have not succeeded to find the
corresponding value in the literature.

\begin{center} (a) Space-time Curvature of the earth
\end{center}
In a first step we want to examine whether the
space-time curvature caused by the
earth may be measurable. Acceleration, rotation, and Riemannian
curvature tensor components are then given by Eqs.
(\ref{accint}), (\ref{rotint}), and (\ref{curvint}) in appendix
C. The three angles $\alpha$, $\beta$, and $\gamma$ describe
the orientation of the interferometer (which is fixed to the
coordinate system as described above, comp. Fig.~1) relative to
the earth. They are defined as follows. In the initial
orientation for which all three angles vanish,
$x^3$ points towards the ceiling of the laboratory, $x^1$ to the
south, and
$x^2$ to the east. We now perform three rotations. The
first is around the $x^3$-axis with the angle $\alpha$ and turns
the apparatus so that the $x^1$-axis does not point
further in north-south-direction. The second is around
the new $x^2$-axis (this is the axis perpendicular to the atom
beam and the laser waves) with angle $\beta$. The third rotation
is around the resulting $x^1$-axis (turning around the incident
atomic beam) with angle $\gamma$ (comp. Fig. 2 for the case
$\alpha =0$).

To get an impression of the order of magnitude of the phase
shifts caused by gravitational acceleration, rotation, and
space-time curvature on earth, we have worked out on the
basis of table I, Eq. (\ref{gesshift}), and Eqs. (\ref{accint})
to (\ref{curvint})
the respective maximal values. The results are listed in
table II. They are obtained in each case for an optimally
adjusted orientation of the interferometer.
Note that for
a maximal influence of $\vec{a}$ another orientation is needed
than for $\vec{\omega}$ or the curvature terms.
In the last column we have set two values in brackets because
they are not observable since the flight time is much longer
than the lifetime of the excited state.

Table II clearly demonstrates that for the two existing
interferometers No.~1 and No.~3 only $\Delta \varphi_a$ is
measurable. The
influence of the earth's rotation is too small and the influence
of the space-time curvature is many orders of magnitudes
too small.

Because of the cubic time dependence of the curvature induced
phase shift this situation changes if the
flight times are in the order of one second.
To enlarge the times of flight $T,T^\prime$ between the lasers
it is necessary to
slow down the atoms and to build a device with larger distances
between the laser beams. Modern laser cooling
techniques allow to build spectrometers
in which the mean velocity of the atoms is as low as 2 m s$^{
-1}$ (see, e.g., Ref. \cite{ertmer92}).
There are some limitations on the magnitude of $T$ and $T^\prime
$. Since in the
interferometer geometry of Fig. 1 leading to $b^{(-1)}$
one part of the atomic beam
is excited between the laser pairs 1-2 and 3-4 the corresponding
time of flight $T$ is limited to be at best of the order of the
lifetime $\tau_l$ of the excited state. If it is substantially
larger the coherence of the atoms will be destroyed by
spontaneous emission happening between the copropagating laser
beams. However, this argument does not hold for the time of
flight $T^\prime$ between the second and
third laser since all atoms moving to the $b^{(-1)}$ output are
unexcited. It is clear that even $T^\prime$ cannot be made
very large since it is very difficult to collimate the
atomic beam over large distances. The loss of atomic flux for a
larger $T^\prime$ leads to a growing integration time for a
given accuracy of the statistics.
Referring to the experiment of Kasevich and Chu
\cite{kasevich91}
where the time of flight is about half a second and the atomic
flux decreases with a factor of about thirty we think that a
total time of flight of one or two seconds should be
possible within the near future. This leads to the
modified specifications No. 2 and No. 4 of the Ramsey device
given in Tab.~I.

Based on these modifications we obtain for the phase shifts the
results given in the rows 2e and 4e of table II. The curvature
induced phase shift $\Delta \varphi_R$ has now become large
enough to be measurable if one manages to separate it from
the shifts $\Delta \varphi_a$ and $\Delta \varphi_{\omega}$
which have become very large and will
contribute to the resulting
shift $\Delta \varphi$ of Eq. (\ref{gesshift}).
This separation could formally be achieved
by laser and atomic beam reversal
\cite{maau92}. This employs the fact that the shifts
(\ref{acc}) to (\ref{aR}) have different powers of $k$ and
$p_1$ and therefore behave differently under beam reversal.
For instance, to isolate $\Delta \varphi_{0103}$ one can
invert the atomic beam
($p_1 \rightarrow -p_1$) and set the device on a turntable so
that $\Delta \varphi_{\omega}\approx 0$, or perform the
experiment at a parallel latitude $\chi = \pi /2 - \vartheta$
where $\omega_2 $ of Eq. (\ref{rotint}) is zero
for certain orientations. In addition, it is necessary to
perform the experiment at least in two different orientations
(by changing the orientation of the lasers)
in order to vary the magnitude of $R_{0103}$ and therefore the
isolated phase shift $\Delta \varphi_{0103}$. Similarly, one
could reverse the laser beams
($k \rightarrow -k$) in order to isolate $\Delta
\varphi_{0303}$. In this case both $T$ and $T^\prime$ have to be
in the order of one second, otherwise $\Delta \varphi_{0303}$
would be too small.

Unfortunately, for all practical realizations
these changes in the orientation are sources of big errors
since the acceleration induced phase shift is much larger than
the curvature shift. According to row 4e of Table II
we have $\Delta \varphi_a \approx 10^6 \Delta \varphi_R$.
In order to get proper results the relative error in the
contribution of the
acceleration has to be of the order of $10^{-6}$. Neglecting
for simplicity the centrifugal force Eq. (\ref{accint})
states that the relevant component of the acceleration is
given by $a_3 \approx \cos \beta \cos \gamma \, 9.81
$m s$^{-2}$.
To estimate the corresponding allowed error $\delta \beta$
in the fixation of the angle $\beta$ we expand $a_3$ around
$\beta =0$. Then the error $\delta
a_3$ is proportional to $\delta \beta^2$ which shows
that $\delta \beta$ has to be smaller than about
$10^{-3}$ rad which is too difficult to manage. Furthermore, for
the orientation $\beta=0$ the component $R_{0103}$ of the
curvature tensor vanishes so that we would loose a part of the
effect. On the other hand, the expansion around $\beta = \pi /4$
for which $R_{0103}$ is
maximal leads to $\delta \beta < 10^{-6}$ rad which is
clearly out of reach.

It should be stressed that the experimental situation is even
worsened by the fact that the atomic beam is bended through
the acceleration of the earth. Thus, to match the atoms, the
laser beams have to be adjusted after each change of an
orientation although the times $T$ and $T^\prime$ can be held
constant by using laser pulses as in Ref. \cite{kasevich91}.

Anandan \cite{anandan83} and Clauser \cite{clauser88} have made
a proposal to eliminate the rotational and accelerational phase
shift, respectively. An
interferometer with crossing beams ("figure eight") is only
sensitive to relative acceleration (curvature).
To obtain this configuration in using four running laser waves
the waves must be travelling into the same direction. In
this case always one of the atomic partial beams must be
excited. We therefore loose the possibility to enlarge the
times of flight $T$ and $T^\prime$.

Accordingly, our final conclusion is that it is not possible to
demonstrate the influence of space-time curvature on Ramsey
interference if the gravitational field of the earth is used.

To close this paragraph we remark that even if both flight times
$T$ and $T^\prime$ are of the order of one second one would
neither be able to detect the
reaction of quantum systems on gravitational radiation
(because of the corresponding $R_{0i0j} \approx 5 \cdot 10^{-28}
m^{-2}$) nor the Lense-Thirring effect, i.e., the dragging
of the inertial reference frame
in free motion caused by the rotation
of the earth (because of the relevant $ \omega_{LT} \approx
5\cdot 10^{-14} s^{-1}$).

\begin{center} (b) The Space-time
Curvature of two lead blocks \end{center}
The problem with the earth gravitational field is that the
influence of curvature, although measurable as far as its
magnitude
is concerned, cannot be separated from the influence of the
acceleration. We therefore have to look for other sources of
gravitational fields where this separation can be performed
easier.
That even much smaller massive bodies such as two lead blocks
can produce in their vicinity a curvature comparable to that of
the earth can be seen by a simple argument. Consider some
spherical body with mass $M=4\pi \rho R^3 /3$, homogeneous
mass density $\rho$, and radius $R$. Its Newtonian potential
in the distance $r$
is proportional to $M/r$. As discussed in Sec. 5 the

acceleration and curvature for an observer at rest relative to
the body are the first and second derivative of this potential,
respectively. Thus, $R_{0l0m} \propto M/r^3$
(here we neglect any angular dependence). In the vicinity
of the body we get $R_{0l0m} \propto \rho$, that is the
curvature is independent of the radius $R$ of the body. This is
not so for the acceleration which is by the same argument
proportional to $R \rho$. It is therefore much larger for the
earth than for small bodies with comparable mass density. Let
us therefore turn to curvature produced by laboratory sized
objects.

In appendix 3 we have derived the curvature components of two
parallel lead blocks (Eqs. (\ref{lbm}) and (\ref{lbm2})) for
the interferometer positioned in the middle between the blocks.
If we adopt the numerical estimation at the end of this
appendix the curvature component $R_{0103}$ has the maximal
value $1.7 \cdot 10^{-22}$ m$^{-2}$ which is larger than the
corresponding component of the earths curvature by a factor of
about 6.6.
It is clear that we can adopt the discussion in paragraph (a)
with few changes. The separation of the shift $\Delta \varphi_R$
produced by the blocks from the shifts caused by
acceleration and rotation of the
earth is not necessary. For a fixed interferometer we can simply
remove the lead blocks thereby changing only $\Delta \varphi_R$.

To estimate the magnitude of $\Delta \varphi_R$
we use the data of
the modified device in row 4 of Tab.~I in Eq. (\ref{gesshift}).
Let us assume an orientation of the interferometer for which
the atoms are initially travelling upwards
orthogonal to the surface of the earth with momentum $p_1$.
This determines the coordinates $x^1 ,x^2 ,x^3$ which are fixed
to the interferometer. Note that the discussion in appendix 3
indicates that
the acceleration $\vec{a}$ which enters into Eq.
(\ref{gesshift})
is the acceleration which the observer needs in order to
avoid the free fall towards the center of the earth, $\vec{a}$
is therefore the negative of the earth's acceleration.
For this arrangement of the interferometer only $R_{0103}$
enters $\Delta \varphi_{aR}$. $\Delta \varphi_{0303}^{(\pm 1)}$
is negligible.
$R_{0103}$ and so roughly $\Delta \varphi_R$
take their maximal values for  $\gamma =0$ and $\beta = \pi /4$.

These angles describe now the orientation of the tilted lead
blocks relative to the interferometer. The initial momentum
$p_1$ (which points to the $x^1$-direction) is not furthermore
parallel to the surfaces of the two blocks, but forms an angle
$\beta$. Taking all this into account we find for the phase
shift the value 0.4 as given in row 4l of Tab.~II.

To sum up: With reference to the gravitational field of two lead
blocks we have shown that the curvature induced shift $\Delta
\varphi_R$
of the Ramsey fringes is within the range of measurability
for experimental set-ups which may be realized in the
near future. They should make it possible
to demonstrate for the first time the influence of
space-time curvature on a quantum system under laboratory
conditions.
\acknowledgements
We thank Prof. Dr. F. Riehle, Dr. C. L\"ammerzahl, T. Pfau, and
C. Kurtsiefer for
valuable discussions. This work was supported by the
Studienstiftung des Deutschen Volkes and by the
commission of the European Community, DG XII.
\appendix{}
We use the conventions of \cite{MTW}, i.e. $ \eta_{\alpha \beta}
= \mbox{diag} (-1,1,1,1)$ and $ R^{\mu}_{\; \; \nu \alpha \beta}

= \Gamma^{\mu}_{\; \; \nu \beta , \alpha} - \cdots $. Latin

indices run from 1 to 3 and tetrad indices are underlined.

We sum over equal indices regardless of their position
(upper or lower). Non-summation over equal indices is denoted
by setting one of them in brackets. The

Tetrads fulfill $ e^{\underline{\alpha}}_{\;\mu}e^{\underline{

\beta}}_{\; \nu}g^{\mu \nu} = \eta^{\underline{\alpha}

\underline{\beta}} $. The Dirac matrices obey the anti

commutator relation $\{ \gamma_{\underline{\alpha}},\gamma_{

\underline{\beta}} \} =2\eta_{\underline{\alpha}\underline{

\beta}}${\bf 1}$_4$. Note that the
sign convention for the metric implies that the $\gamma_{
\underline{\mu}}$ matrices have an additional factor of $i$
compared to the matrices used in particle physics.
We define $\alpha_{\underline{i}} \equiv \gamma_{\underline{0}}

\gamma_{\underline{i}}$, $\gamma_{\underline{5}} \equiv

i\gamma^{\underline{0}}\gamma^{\underline{1}}\gamma^{
\underline{2}}\gamma^{\underline{3}}$, $\Sigma_{\underline{k}}
\equiv -i \varepsilon_{ijk} \gamma_{\underline{i}}

\gamma_{\underline{j}} /2$. The spinor connection is given by
\beq \Gamma_{\mu} =-{\st \frac{1}{4}}\gamma_{\underline{\alpha}}

     \gamma_{\underline{\beta}}\; e^{\underline{\alpha}

     \nu}\nabla_{\mu} e^{\underline{\beta}}_{\; \, \nu} \; ,\eeq
where $\nabla_{\mu}$ is the covariant derivative acting on
vectors.

The hypersurface independent scalar product between spinors is

\cite{parker80}
\beq (\phi ,\psi ) = - \int d^3 x \sqrt{-g}\; \phi^+

     \gamma^{\underline{0}}\; e^{0}_{\; \; \underline{\mu}}\;

     \gamma^{\underline{\mu}}\; \psi \label{sp} \eeq
The contravariant components of the metric in Fermi coordinates
are up to order $O((x^l)^2)$ (comp. (\ref{metrik}))
\bea g^{00} &=& -1 + 2 (\vec{a}\cdot \vec{x}) - 3(\vec{a}\cdot

     \vec{x})^2 + R_{0k0l} x^k x^l  \nn \\
     g^{0i} &=& (1-2 \vec{a} \cdot \vec{x}) (\vec{\omega}\times

     \vec{x})^i -{\st \frac{2}{3}} R_{0lim} x^l x^m  \nn \\
     g^{ij} &=& \delta^{ij} + {\st \frac{1}{3}} R_{iljm}x^l x^m

     - (\vec{\omega}\times \vec{x})^i (\vec{\omega}\times

     \vec{x})^j
\eea

After some lengthy algebra we find for the Christoffels (to

$O(x^l )$)
\bea \Gamma^0_{\; \, 00} &=& \dot{\vec{a}}\cdot \vec{x} +

     \vec{a}\cdot (\vec{\omega}\times \vec{x}) \nn \\
     \Gamma^i_{\; \, 00} &=& R_{0i0l}x^l +(1+\vec{a}\cdot

     \vec{x})a_i +(\dot{\vec{\omega}}\times \vec{x})_i
     -((\vec{\omega}\times \vec{x}) \times \vec{\omega})_i \nn

     \\ \Gamma^0_{\; \, 0i} &=& R_{0i0l}x^l +(1-\vec{a}\cdot

     \vec{x})a_i \nn \\
     \Gamma^j_{\; \, 0i} &=& R_{0mij}x^m

     -\varepsilon_{jil}\omega^l -a_i (\vec{\omega}\times

     \vec{x})_j \nn \\ \Gamma^0_{\; \, ij} &=& -{\st

     \frac{1}{3}}(R_{0imj} +R_{0jmi})x^m  \nn \\
     \Gamma^k_{\; \, ij} &=& -{\st \frac{1}{3}}(R_{kijm}

     +R_{kjim})x^m
\eea
and for the tetrads
\bea e^{\underline{0}}_{\; \, 0} &=& 1+(\vec{a}\cdot \vec{x})

     -{\st \frac{1}{2}} R^0_{\; \,l0m}x^l x^m  \nn \\
     e^{\underline{j}}_{\; \, 0} &=& -{\st \frac{1}{2}}R^j_{\;

     \,l0m} x^l x^m + \varepsilon_{jlm}\omega^l x^m \nn \\
     e^{\underline{0}}_{\; \, i} &=& -{\st \frac{1}{6}}R^0_{\;

     \, lim}x^l x^m  \nn \\
     e^{\underline{j}}_{\; \, i} &=& \delta^j_{\; \, i} - {\st

     \frac{1}{6}}R^j_{\; \,lim}x^l x^m
\label{tetrad} \eea
The spinor connections in Fermi coordinates are
\bea \Gamma_0 &=& {\st \frac{1}{2}}

     \gamma_{\underline{0}}\gamma_{\underline{i}}
     (R_{i00m}x^m -a_i) +{\st \frac{1}{4}}\gamma_{\underline{i}}
     \gamma_{\underline{j}}(R_{ij0m}x^m +\varepsilon_{ijl}

     \omega_l )\nn \\ \Gamma_i &=& {\st \frac{1}{4}}\gamma_{

     \underline{0}}\gamma_{\underline{j}} R_{0jmi}x^m + {\st

     \frac{1}{8}}\gamma_{\underline{j}}\gamma_{\underline{k}}

     R_{jkim}x^m
\eea
and the scalar product is given by
\beq (\phi ,\psi ) = \int d^3 x \phi^+ [1-{\st \frac{1}{6}}

     R_{ilim}x^l x^m -{\st \frac{1}{6}} R_{0lim}x^l x^m

     \alpha_{\underline{i}}]\psi \label{skalprod} \eeq
\appendix{}
Here we generalize the calculation of the corrections to the
Coulomb potential in Fermi coordinates given by
Parker \cite{parker80} to the case of an accelerating and
rotating observer.

With reference to the vector potential $A_{\mu}$ the
covariant Maxwell equations in curved spacetime are given by
\beq \nabla^{\lambda}\nabla_{\lambda}A_{\mu}-R_{\mu}^{\; \,

     \nu}A_{\nu} = -4\pi j_{\mu} \label{maxwell} \eeq
where we have used the Lorentz gauge $\nabla_{\mu} A^{\mu} =0$.

In order to determine the

influence of non-inertial motion and spacetime curvature on
the electromagnetic field of a point charge we write the

Maxwell equations in adapted Fermi coordinates.
Since the charge is now resting at the origin of the coordinate
system the current is given by
\beq j_{\mu} = -Ze \delta (\vec{x}) \; (1,0,0,0) \label{strom}
\eeq

Inserting the metric (\ref{metrik}) and the current

(\ref{strom}) into Eq. (\ref{maxwell}) we arrive at

\bea & &  [ \delta^{ij} +{\st \frac{1}{3}}R_{iljm}x^l x^m -
     (\vec{\omega} \times \vec{x})_i (\vec{\omega} \times

     \vec{x})_j ]  \partial_i \partial_j A_0 +

      A_{0,i}[ -{\st \frac{2}{3}}R_{im}x^m -{\st \frac{5}{
      3}}R_{0i0m}x^m +(\vec{a}\cdot \vec{x} -1)a_i \nn \\
      & & +((\vec{\omega} \times \vec{x})\times \vec{\omega})_i

      ] + A_{i,j}[ 2R_{0mij}x^m +2((\vec{\omega} \times

      \vec{x})_i a_j - (\vec{\omega} \times \vec{x})_j a_i )+2

      \varepsilon_{ijl}\omega_l ] \nn \\
      & & = 4\pi Z e \delta (\vec{x})
\eea
and
\bea & & [ \delta^{ij} +{\st \frac{1}{3}}R_{iljm}x^l x^m -
     (\vec{\omega} \times \vec{x})_i (\vec{\omega} \times

     \vec{x})_j ] \partial_i \partial_j A_k +
     A_{0,j}[ {\st \frac{2}{3}}(R_{0kmj}+R_{0jmk})x^m -2

     (\vec{\omega} \times \vec{x})_j a_k ] \nn \\

     & & + A_{j,i}[ {\st \frac{2}{3}}(R_{jikm}+R_{jkim})x^m +2
     \varepsilon_{jkl} \omega_l (\vec{\omega} \times \vec{x})_i

     ] + A_{k,j}[ {\st \frac{1}{3}}(R_{0j0m}-2R_{jm})x^m

     +(1-\vec{a}\cdot \vec{x})a_j \nn \\
     & & +((\vec{\omega} \times \vec{x})\times \vec{\omega})_j ]

     -{\st \frac{2}{3}}R_k^{\; \, \lambda}A_{\lambda} +{\st

     \frac{1}{3}}R_{0k0j}A_j +(\vec{\omega}\times \vec{a})_k

     A_0 -a_k (\vec{a} \cdot \vec{A}) +(\vec{\omega}\times

     (\vec{A}\times \vec{\omega}))_k \nn \\ & & = 0

\eea
All equations are understood to be correct only up to second
order in the spatial coordinates $x^l$, derivatives of second
order equations are of first order and so on.

We now divide the electromagnetic potential into two parts,
one unperturbed, which is simply the Coulomb potential in flat
spacetime, and one, $A_{\mu}^{(1)}$, which

contains the corrections up to order $O(x^l )$:
\beq A_0 = -\frac{Ze}{r} +A_0^{(1)}\; , \quad A_i = A_i^{(1)}
     \label{CoulAnsatz}\; .\eeq
The 2nd derivatives of $A_{\mu}^{(1)}$ are then of the order

$O((x^l )^{-1})$ so that we can drop all terms which are of
higher order in the Maxwell equations. This leads to
\bea \partial_i \partial_i A_0^{(1)} - \vec{a} \cdot

     \vec{\nabla} A_0^{(1)} +2 \varepsilon_{ijl} \omega_l

     A_{i,j}^{(1)} & = & \frac{Ze}{r^3} \{ \vec{a}

     \cdot \vec{x} (1-\vec{a} \cdot \vec{x}) + (\vec{\omega}

     \times \vec{x})^2 \nn \\
     & & +{\st \frac{1}{3}}(R_{lm} +4R_{0l0m})x^l x^m \}
     \nn \\
     \partial_i \partial_i A_k^{(1)} +a_j A_{k,j}^{(1)} &=&

     \frac{Ze}{r^3}\{ -{\st \frac{2}{3}}R_{0jmk}x^m x^j +{\st

     \frac{2}{3}}R_{k0} r^2 +(\vec{\omega} \times \vec{a})_k

     r^2 \} \label{ff} \eea



The curvature terms of Eq. (\ref{ff}) are in

disagreement with Eq. (7.9) of Ref. \cite{parker80}.


The solution of these equations is found to be
\bea A_0^{(1)} & = & -\frac{Ze}{2}\frac{\vec{a}\cdot \vec{x}}{r}

     +Ze \{ {\st \frac{1}{4}} \vec{\omega}^2 -{\st

     \frac{3}{8}}\vec{a}^2 + {\st \frac{1}{12}}(R+5R_{00})\} r

     \nn \\ & & +Ze \{ {\st \frac{1}{4}}\omega_l \omega_m +{\st

     \frac{1}{8}}a_l a_m -{\st \frac{1}{12}}(R_{lm}+4R_{0l0m})\}

     \frac{x^l x^m}{r} \nn \\
     A_k^{(1)} &=& \frac{Ze}{2}(R_{0k} +(\vec{\omega}\times

     \vec{a})_k )r + \frac{Ze}{6}R_{0lmk}\frac{x^l x^m}{r}

     \label{mixing} \eea


which satisfies the Lorentz condition. Again, the curvature
terms are slightly different from those given in Ref.

\cite{parker80}.
\appendix{}
In this appendix we scetch the derivation of the components of
the Riemannian curvature tensor in Fermi coordinates for
the rotating earth, modeled by the Schwarzschild metric, as seen
by an observer resting on it and for two lead blocks in the
limit of weak gravity.

The first case was already studied by Parker and Pimentel
\cite{parker82} for an observer in radial and circular
geodesic motion. We will modify their results for the
worldline $P_0 (\tau)$ given by

$\dot{r} =\dot{\vartheta}= 0$ and $\dot{\varphi} :=
\tilde{\omega}$ with $\tilde{\omega} =2\pi$ day$^{-1}$
given by the rotation of the earth. The dot
denotes the derivative with respect to the observers
proper time $\tau$.
Let $R_s = 2 G M /c^2$ be the Schwarzschild radius of the
central body where $M$ is its mass and
$c$ is reinserted ($R_s \approx 8.9$ mm for the earth).
$G$ is Newtons constant. Defining
$X \equiv 1-R_s /r$ the components of the curvature tensor
in standard Schwarzschild coordinates (see, e.g., Ref.
\cite{parker82}) are given by
\beq \begin{array}{lclcl} R_{rtrt} = -{\dst R_s \over \dst r^3}

     &,& R_{\vartheta t \vartheta t}= {\dst R_s X\over \dst 2r}
     &,& R_{\varphi t \varphi t}= {\dst
     R_s  X\over \dst 2r} \sin^2 \vartheta \\

     R_{r \vartheta r \vartheta}= -{\dst R_s\over \dst 2rX} &,&
     R_{\vartheta \varphi \vartheta \varphi}= R_s r
     \sin^2 \vartheta  &,& R_{r \varphi r \varphi} = -{\dst
     R_s\over \dst 2rX} \sin^2 \vartheta \end{array} \eeq
The acceleration $a^\mu$ of the observer can be read off from
its equation of motion
\beq \ddot{x}^{\mu} + \Gamma^{\mu}_{\; \nu \lambda}\dot{x}^{\nu}

     \dot{x}^{\lambda} = a^{\mu} \label{ngeod} \eeq
which describes the worldline $P_0 (\tau )$ with tangent vector
$u^\mu = \dot{x}^\mu$. Taking into account the conditions $u^\mu
u_{\mu}=-1$ and $u^\mu a_{\mu}=0$, we find
\beq a^r = \frac{R_s}{2 r^2} - \tilde{\omega}^2 r (1-
     \frac{3 R_s}{2r})\sin^2 \vartheta \; , \quad a^\vartheta =
     - \tilde{\omega}^2 \sin \vartheta \cos \vartheta \eeq
We observe that the first part of $a^r$ is the negative of

Newtons acceleration. This is reasonable with regard to the
physical meaning of $a^\mu$ since the
surface of the earth prevents the observer to fall freely.
The second part and $a^\vartheta$ represent the centrifugal
force. Note that in $a^r$ a general
relativistic correction $-3 R_s /2$
is present. This is negligible for the earth but is of interest
in the case of a black hole where for $r < 1.5 R_s$ the
radial component of the centrifugal force changes the direction
(see also Ref. \cite{fliehkraft}).

We assume the comoving tetrad ($e_{\underline{0}^\prime}^\alpha
= u^\alpha$) to be fixed to the rotating earth according to
$e_{\underline{1}^\prime} \sim \partial_{\vartheta}$ and
$e_{\underline{3}^\prime} \sim \partial_{r}$ (dashed lines in
Fig. 2). With $e_{\underline{0}^\prime} \propto \partial_{\tau}$

and $e_{\underline{\alpha}^\prime} \cdot e_{\underline{\beta}
^\prime} = \eta_{
\underline{\alpha}^\prime \underline{\beta}^\prime} $ it is

found to be
\bea e_{\underline{0}^\prime} = \frac{W}{\sqrt{X}} \partial_t
     + \tilde{\omega} \partial_{\varphi} & ,& e_{\underline{1
     }^\prime} = r^{-1} \partial_{\vartheta} \nn \\
     e_{\underline{2}^\prime} = \frac{\tilde{\omega} r \sin
     \vartheta}{\sqrt{X}} \partial_t + \frac{W}{r \sin
     \vartheta} \partial_{\varphi} &,&
     e_{\underline{3}^\prime} = \sqrt{X} \partial_r

     \label{schwtet} \eea
whereby $W \equiv \sqrt{1 + r^2 \tilde{\omega}^2 \sin^2
\vartheta}$. The rotation $\omega^\alpha$ of the
tetrad represents physically the rotation relative to a
Fermi transported tetrad which may be fixed by gyroscopes.
It is this rotation which enters the metric in Fermi coordinates
as in (\ref{metrik}). Mathematically $\omega^\alpha$ is given by
(comp. Ref. \cite{MTW})
\beq \frac{\dst D e_{\underline{\alpha}^\prime}^\mu}{D \tau} = -

      [a^{\mu} u^{\nu} - a^{\nu} u^{\mu}

     + u_{\alpha} \omega_{\beta} \varepsilon^{\alpha \beta \mu

     \nu}] e_{\underline{\alpha}^\prime \nu} \eeq
Inserting (\ref{schwtet}) we find
\beq \omega_r = \frac{W \tilde{\omega} \cos \vartheta}{\sqrt{X}}
     \; , \quad \omega_{\vartheta} = -\frac{W \tilde{\omega} r
     \sin \vartheta}{\sqrt{X}} (1-{3R_s \over 2r}) \eeq
Note again the general relativistic corrections proportional
to $R_s /r$.

Still referring to the tetrad (\ref{schwtet}) we work out the
components
of $a_{\mu}$, $\omega_{\mu}$, and $R_{\mu \nu \rho \sigma}$ on
the worldline $P_0 (\tau )$ in the respective Fermi coordinates
adjusted to the tetrad. This amounts to a projection with the
tetrad. Using
\beq R_{\underline{\alpha}^\prime\underline{\beta}^\prime
     \underline{\gamma}^\prime
     \underline{\delta}^\prime} = R_{\mu \nu \rho \sigma}

     e_{\underline{\alpha}^\prime}^{\mu}

     e_{\underline{\beta}^\prime}^{\nu}
     e_{\underline{\gamma}^\prime}^{\rho}

     e_{\underline{\delta}^\prime}^{\sigma} \eeq
we find for the curvature in the Fermi coordinates of the
observer
\beq \begin{array}{ll}

     R_{\underline{0}^\prime \underline{1}^\prime
     \underline{0}^\prime \underline{1}^\prime} = -
     R_{\underline{2}^\prime \underline{3}^\prime
     \underline{2}^\prime \underline{3}^\prime} =
     \frac{\dst R_s}{\dst 2 r^3} \{ 1+3 r^2 \tilde{\omega}^2
     \sin^2 \vartheta \} \; ,&
     R_{\underline{0}^\prime \underline{2}^\prime
     \underline{0}^\prime \underline{2}^\prime} = -
     R_{\underline{1}^\prime \underline{3}^\prime
     \underline{1}^\prime \underline{3}^\prime} =
     \frac{\dst R_s}{\dst 2 r^3} \nn \\[3mm]
      R_{\underline{0}^\prime \underline{3}^\prime
     \underline{0}^\prime \underline{3}^\prime} = -
     R_{\underline{1}^\prime \underline{2}^\prime
     \underline{1}^\prime \underline{2}^\prime} = -
     \frac{\dst R_s}{\dst r^3} \{ 1+{3\over 2} r^2
     \tilde{\omega}^2 \sin^2 \vartheta \} \; ,&
     R_{\underline{0}^\prime \underline{1}^\prime
     \underline{2}^\prime \underline{1}^\prime} = -
     R_{\underline{0}^\prime \underline{3}^\prime
     \underline{2}^\prime \underline{3}^\prime} =
     \frac{\dst 3 R_s}{\dst 2 r^2} W \tilde{\omega} \sin
     \vartheta  \end{array} \label{curv1} \eeq
For the freely falling observer see Ref. \cite{parker82}.
The case of gravitational waves is discussed in Ref.
\cite{leen83}.
Correspondingly we obtain for the acceleration
\beq a_{\underline{1}^\prime} =  - r \tilde{
     \omega}^2 \sin \vartheta \cos \vartheta \; , \quad
     a_{\underline{3}^\prime} = \frac{R_s}{2 r^2\sqrt{X}} -
     {\tilde{\omega}^2 \over \sqrt{X}} (r- \frac{3}{2}
     R_s)\sin^2 \vartheta \eeq
and for the rotation
\beq \omega_{\underline{1}^\prime}= -\frac{W \tilde{\omega}
     \sin \vartheta}{\sqrt{X}}\left (1-{3 R_s \over 2r}\right
     )\; , \quad \omega_{\underline{3}^\prime} = W
     \tilde{\omega} \cos \vartheta \eeq

To describe  arbitrary orientations of the experimental set-up
'fixed` to the tetrad, we go over to a different orientation
of the three vectors $e_{\underline{i}^\prime} \rightarrow
e_{\underline{i}}$ by means of a Lorentz transformation. Vector
components change according to
\beq e_{\underline{\alpha}} = \Lambda_{\underline{\alpha}
     }^{\; \;\underline{\beta}^\prime}\: e_{\underline{\beta}
     ^\prime}\; , \quad R_{\underline{\alpha \beta

     \gamma \delta}} = \Lambda_{\underline{\alpha}
     }^{\; \;\underline{\mu}^\prime}\: \Lambda_{\underline{

     \beta}}^{\; \;\underline{\nu}^\prime}\:

     \Lambda_{\underline{ \gamma}}^{\; \;\underline{
     \rho}^\prime} \: \Lambda_{\underline{\delta}}^{\;

     \;\underline{\sigma}^\prime}R_{\underline{\mu^\prime \nu
     ^\prime \rho^\prime
     \sigma^\prime}} \label{drehung1} \eeq
In detail we perform three rotations:
the first around the $x^3$ axis with angle $\alpha$
(turning the apparatus on the earth's surface), the second around

the {\em new} $x^2$ axis with angle $\beta$
(the axis perpendicular

the the atom beam and the laser waves), and the third around
the {\em new} $x^1$ axis with angle $\gamma$ (turning around the

atomic beam). This results in a matrix
\beq \Lambda_{\underline{\alpha}}^{\; \;
     \underline{\beta}^\prime}
     = \left ( \begin{array}{cccc} 1 & 0 & 0 & 0 \\ 0 & \cos
     \alpha \cos \beta & \sin \alpha \cos \beta & -\sin \beta \\
     0 & -\sin \alpha \cos \gamma + \cos \alpha \sin \beta \sin
     \gamma & \cos \alpha \cos \gamma + \sin \alpha \sin \beta
     \sin \gamma & \cos \beta \sin \gamma \\ 0 & \sin \alpha
     \sin \gamma + \cos \alpha \sin \beta \cos \gamma & -\cos
     \alpha \sin \gamma + \sin \alpha \sin \beta \cos \gamma &
     \cos \beta \cos \gamma \end{array} \right )

     \label{drehung2} \eeq

The relevant components of the various physical quantities
which enter into
the phase shift (\ref{gesshift}) are now given by
\bea a_{\underline{3}} &=& \frac{R_s}{2r^2 \sqrt{X}} \cos \beta
     \cos \gamma \; -\tilde{\omega}^2 \sin \vartheta \Bigg \{ r
     \cos \vartheta [\sin \alpha \sin \gamma + \cos \alpha \sin
     \beta \cos \gamma ] \nn \\ & &
     + \frac{r - 3R_s/2}{\sqrt{X}} \sin
     \vartheta \cos \beta \cos \gamma \Bigg \}

     \label{accint} \eea
\beq \omega_{\underline{2}} = W \tilde{\omega} \Bigg \{ \cos
     \vartheta \cos \beta \sin \gamma + \frac{2 r -3R_s}{2r
     \sqrt{X}} \sin \vartheta [\sin \alpha \cos \gamma - \cos
     \alpha \sin \beta \sin \gamma ] \Bigg \}

     \label{rotint} \eeq
\bea R_{\underline{0103}} &=& \frac{3R_s}{2r^3} \cos \beta \, \{
     \sin \beta \cos \gamma + r^2 \tilde{\omega}^2 \sin^2
     \vartheta \, [ \sin \alpha \cos \alpha \sin \gamma +
     (1+\cos^2 \alpha )\sin \beta \cos \gamma ] \} \nn \\
     R_{\underline{0303}} &=& \frac{R_s}{2r^3} \Big \{ 1 - 3
     W^2 \cos^2 \beta \cos^2 \gamma + 3r^2 \tilde{\omega}^2
     \sin^2 \vartheta \, (\sin \alpha \sin \gamma + \cos \alpha
     \sin \beta \cos \gamma )^2 \Big \}
     \label{curvint} \eea
Some comments are necessary in order to connect these equations
with Sec. 7.

In Fermi coordinates acceleration, rotation, and curvature are
taken with respect to the tetrad of the observer
on $P_0(\tau )$. Therefore, we
can take the results of this appendix over to the main part of
the paper simply by taking not underlined (Fermi coordinate)
indices instead of underlined (tetrad) ones.

$r$ becomes the radius $R_\oplus \approx 6378$ km of the earth.

We now turn to the calculation of the space-time curvature
of two parallel lead blocks. This can be done as described in
Sec. 6 by using Eqs. (\ref{naehe}) and (\ref{krupot}). For the
two lead blocks the components of the energy momentum tensor
are vanishing except for $T_{00}$ which we assume to have the
value $\rho$ in the region $a<|x^3|<b$, $|x^1|<l_1$,
$|x^2|<l_2$. Here $\rho$ is the mass density of lead.
Inserting this in Eq. (\ref{krupot}) and introducing the
functions
\bea g(u,v,w) &:=& \mbox{sgn}(uvw) \arcsin \left [ \frac{(w^2
     -u^2)(v^2+u^2)-2u^2w^2}{(v^2+u^2)(w^2+u^2)}\right ] \nn \\
     h(w,a) &:=& \frac{1}{a}

     \mbox{arcsinh} \left ( \frac{w}{a}
     \right ) \eea
and the abbreviations $l_1^\pm := \pm l_1 -x^1$,
$l_2^\pm := \pm l_2 -x^2$, and $w^\pm_q
:= \sqrt{(l_1^\pm)^2 + (x^3+q)^2}$ as well as
$a_\pm := a \pm x^3$ and $b_\pm := b\pm x^3$
we find for the relevant components
\bea R_{0103} &=& \frac{- \kappa \rho}{8\pi} \bigg \{
      h(l_2^+,w^+_a) - h(l_2^-,w^+_a)
     -h(l_2^+,w^-_a) + h(l_2^-,w^-_a)
     -h(l_2^+,w^+_b) + h(l_2^-,w^+_b) \nn \\ & & \hspace{5mm}
     +h(l_2^+,w^-_b) - h(l_2^-,w^-_b)

     +h(l_2^+,w^+_{-b}) - h(l_2^-,w^+_{-b})
     -h(l_2^+,w^-_{-b})+h(l_2^-,w^-_{-b})\nn \\ & &\hspace{5mm}
     -h(l_2^+,w^+_{-a}) + h(l_2^-,w^+_{-a})
     +h(l_2^+,w^-_{-a}) - h(l_2^-,w^-_{-a}) \bigg \}
     \label{lb1} \eea
and
\bea R_{0303} &=& \frac{\kappa \rho}{16\pi} \bigg \{
     -g(a_+,l_1^+,l_2^+) + g(a_+,l_1^-,l_2^+)
     +g(a_+,l_1^+,l_2^-) - g(a_+,l_1^-,l_2^-)\nn \\ & &
     \hspace{12mm}
     +g(b_+,l_1^+,l_2^+) - g(b_+,l_1^-,l_2^+)
     -g(b_+,l_1^+,l_2^-) + g(b_+,l_1^-,l_2^-)\nn \\ & &
     \hspace{12mm}
     +g(b_-,l_1^+,l_2^+) - g(b_-,l_1^-,l_2^+)
     -g(b_-,l_1^+,l_2^-) + g(b_-,l_1^-,l_2^-)\nn \\ & &
     \hspace{12mm}
     -g(a_-,l_1^+,l_2^+) + g(a_-,l_1^-,l_2^+)
     +g(a_-,l_1^+,l_2^-) - g(a_-,l_1^-,l_2^-)
      \bigg \} \label{lb2} \eea
In principle we need the curvature components only at the
position of the observer who is assumed to rest in the middle
between the plates. But it is instructive to test with the aid
of Eqs. (\ref{lb1}) and (\ref{lb2}) whether the curvature
components are roughly constant inside the interferometer.
If this is not the case the expansion (\ref{metrik}) of the
metric would be insufficient to describe the physical
situation.

In the middle between the plates ($\vec{x}=0$) the curvature
components are

\bea R_{0101}&=& \kappa \rho (g(l_1,l_2,b)-g(l_1,l_2,a))\nn \\
     R_{0202}&=& \kappa \rho (g(l_2,l_1,b)-g(l_2,l_1,a))\nn \\
     R_{0303}&=& \kappa \rho (g(b,l_1,l_2)-g(a,l_1,l_2))\; ,
     \label{lbm} \eea
all other
$R_{0l0m}$ are zero. It is not difficult to show that for an
observer resting in a static gravitational field of the type
(\ref{naehe}) the components of any tensor linear in
$h_{\mu \nu}$ taken on the
worldline are the same for the coordinate system in which the
gravitational field does not depend on the time (as used in
Eq. (\ref{lbm})) and the Fermi coordinates of the observer.
Eq. (\ref{lbm}) is therefore equivalent to Eq. (\ref{curv1})
in the case of the earth (Simply underline the indices and add
a prime). As before we have the freedom to
rotate the frame of reference along the lines of Eqs.
(\ref{drehung1}) and (\ref{drehung2}) in such a way that the
$R_{\underline{0103}}$ component becomes different from zero. If
it happens that $R_{\underline{0^\prime 1^\prime 0^\prime
1^\prime}}$ equals $R_{\underline{0^\prime 2^\prime
0^\prime 2^\prime}}$ as it is the case whenever $l_1$ and $l_2$
are equal then the relevant components of the curvature tensor
are
\bea R_{\underline{0103}} &=& (R_{\underline{0^\prime 1^\prime
     0^\prime 1^\prime}} - R_{\underline{0^\prime 3^\prime
     0^\prime 3^\prime}}) \sin \beta \cos \beta \cos \gamma
     \nn \\
     R_{\underline{0303}} &=& R_{\underline{0^\prime 1^\prime
     0^\prime 1^\prime}} (\sin^2 \gamma + \sin^2 \beta \cos^2
     \gamma ) + R_{\underline{0^\prime 3^\prime
     0^\prime 3^\prime}} \cos^2 \beta \cos^2 \gamma
     \label{lbm2} \eea
where the primed components are the same as in Eq. (\ref{lbm}).
Just to put in some numbers, for $a:b:l_1:
l_2 = 0.1:1:2:2$ we have

$R_{\underline{0^\prime 1^\prime 0^\prime 1^\prime}} =
R_{\underline{0^\prime 2^\prime 0^\prime 2^\prime}}
\approx 0.57 \kappa \rho$ and $R_{\underline{0^\prime 3^\prime
0^\prime 3^\prime}}\approx -1.15 \kappa \rho$. For lead we
have (in SI units) $\rho \approx 1.1 \cdot 10^4 $ kg m$^{-3}$
and therefore $\kappa \rho \approx 2\cdot 10^{-22}$ m$^{-2}$.
The maximal value of $R_{\underline{0103}}$ is therefore
$0.86 \kappa \rho$ for $\beta = \pi /4$ and $\gamma =0$.
The curvature in the middle between the blocks is therefore
larger than the curvature of the earth on its surface by a
factor of about six.

We have to remark that we have chosen an arrangement of two
identical lead blocks since from
symmetry arguments one can see that the acceleration in the
middle between the blocks is zero so that in this frame of
reference there is no change in the acceleration if we remove
the two blocks.

\figure{The laser waves in the Ramsey experiment split the
atomic beam. The dashed lines represent atoms in the excited
state $b$. The four parts of the beam shown here build up two
interferometer geometries.\label{fig1}}
\figure{The orientation of the interferometer on the earth
after the performance of three rotations. $\chi = \pi /2 -
\vartheta $ is the
parallel latitude of the laboratory. The atomic beam is
parallel to $x^1$, the lasers are parallel to
$x^3$.\label{fig2}}
\begin{table}
\caption{Lifetime, wavelength, atomic mass, atomic mean
velocity, and the distances between the lasers of two existing
and two hypothetical Ramsey devices.} \label{tab1}
\begin{tabular}{||c|l||c|c|c|c|c|c||}
  No. & Reference & $\tau_l$ [ms] & $\lambda$ [nm] & $m$ [kg] &

  $v_1$ [m/s] & $T$ [ms]& $T^\prime$ [ms]

  \\ \hline \hline
1 & \cite{riehle91} & 0.4 & 657 & $6.7 \cdot 10^{-26}$
   & 700 & $1.86\cdot 10^{-2}$ & $4.7\cdot 10^{-2}$ \\ \hline
2 & \cite{riehle91} modified & 0.4 & 657 & $6.7 \cdot 10^{-26}$
   & 2 & 0.2 & 1000 \\ \hline
3 & \cite{ertmer92} & 4.6 & 457 & $4 \cdot 10^{-26}$
   & 700 & $1.7\cdot 10^{-2}$ & $6 \cdot 10^{-2}$ (?) \\ \hline
4 & \cite{ertmer92} modified & 4.6 & 457 & $4 \cdot 10^{-26}$
   & 2 & 3 & 1000 \\ 

\end{tabular}
\end{table}
\begin{table}
\caption{The calculated phase shift induced by acceleration,
rotation, and space-time curvature for the experimental data
given in Tab.~I. "e" stands for earth curvature, "l" for two
lead blocks. The orientation of the interferometer with
respect to the earth or the lead blocks
is given by the angels $\alpha ,\beta
,\gamma$, and $\vartheta$. See text and Fig.\ 2 for
details.}\label{tab2}
\begin{tabular}{||c|l||c|c|c||}
  No.& Ref. & $\Delta \varphi_a $ & $\Delta \varphi_{\omega}$ &
  $\Delta \varphi_R$ \\ \hline

\multicolumn{2}{||c||}{orientation} & $\beta ,\gamma =0$ & $
  \gamma =0,\alpha =\vartheta =\pi /2$ & $\beta =\pi /4 ,\gamma
  =0$ \\ \hline  \hline
1e & \cite{riehle91} & $-0.11$ & $1.9  \cdot 10^{-4}$ & $-8
   \cdot 10^{-10}$ \\ \hline
2e & \cite{riehle91} mod.  & $-1.9 \cdot 10^4$ & $0.55$&
  $-0.035$ \\ \hline
3e & \cite{ertmer92} & $-0.17$ & $2.8 \cdot 10^{-4}$ &

  $-1.4 \cdot 10^{-9}$ \\ \hline
4e & \cite{ertmer92} mod. & $-4 \cdot 10^5$ & $11.8$ & $-0.25$

   \\ \hline \hline
4l & Lead, \cite{ertmer92} mod. & -- & -- & $0.4$  \\

\end{tabular}
\end{table}
\end{document}